\documentclass{PoS}

\title{New Experiments to Measure the Muon Anomalous Gyromagnetic Moment}

\ShortTitle{New Muon $g-2$ Experiments}

\author{\speaker{Michael Eads}\thanks{on behalf of the New Muon ($g-2$) collaboration at Fermilab and the E-34 collaboration at J-PARC}\\
        Northern Illinois University\\
        E-mail: \email{eads@fnal.gov}}

\abstract{The magnetic moment is a fundamental property of particles. The measurement of these magnetic moments and the comparison with the values predicted by the standard model of particle physics is a way to test our understanding of the fundamental building blocks of our world. In some cases, such as for the electron, this comparison has resulted in confirmation of the standard model with incredible precision. In contrast, the magnetic moment of the muon has shown a long-standing disagreement in the measured and the predicted value. There is currently a tantalizing three-standard-deviation difference between the current best measurement (with a precision of 0.54 ppm) and the state-of-the-art standard model prediction. This represents one of the very few experimental hints for physics beyond the standard model. There are currently two major experimental efforts underway to improve the precision of the muon magnetic moment measurement. The first is an evolution of the E-821 experiment, originally located at Brookhaven National Laboratory in the United States. This is experiment, E-989, is located at Fermilab and will measure the spin precession rate of positive muons in a 14-m diameter storage ring using decay positrons. The goal of the experiment is to reduce the current experimental uncertainty by a factor of three. The experiment is currently being constructed and aims to start taking data in 2017. An alternative, and very complementary, experiment is being planned at J-PARC in Japan. This experiment, E-34, will utilize low energy, ultra-cold muons in a much smaller storage ring. This experiment aims for a similar precision to the Fermilab experiment and aims to begin data taking on a similar timescale.}

\FullConference{Flavor Physics \& CP Violation 2015\\ 
May 25-29, 2015\\ 
Nagoya, Japan}

\begin{document}

\section{Introduction}

The goal of particle physics is to understand the world around us at the most fundamental level. That is, we wish to discover the fundamental particles which comprise the universe and to understand the interactions between these fundamental particles. One way to accomplish this goal is through the precision measurement of fundamental particle properties. 

One such particle property is a particle's magnetic moment, which is a measure of how strongly a particle will interact with a magnetic field. For a point particle of unit charge, the magnetic moment is given by
\begin{equation}
\overrightarrow{\mu} = g \frac{e}{2m} \overrightarrow{s}
\end{equation}
where $e$ is the electron charge, $m$ is the mass of the particle, and $s$ is the spin of the particle. $g$ is a proportionality factor and is expected to slightly larger than two for spin-1/2 particles. For this reason, experimental measurements are often quoted as $(g-2)$ or $a = (g-2)/2$, known as the anomalous magnetic moment. Measurements of $a$ offer a unique opportunity to test the standard model (SM) of particle physics, as this quantity can be experimentally measured and can also be calculated in the SM to great precision. New particles or interactions could change $a$, meaning that these measurements are a promising place to look for new physics. For electrons, the agreement between measurement \cite{electron_measurement} and prediction \cite{electron_theory} is at the part-per-trillion level. The next logical step is to measure the muon anomalous magnetic moment ($a_\mu$), which could have enhanced sensitivity to new physics due to the larger mass of the muon. 

The measurement of the muon anomalous magnetic moment has a long history, highlighting the interplay between experiment and theory \cite{muon_review}. The current prediction for $a_\mu$ has been calculated to five loops in Quantum Electrodynamics \cite{theory_prediction}. The uncertainty of 0.4 parts-per-million (ppm) is dominated by the hadronic contributions. The current best measurement of $a_\mu$ is from the E821 experiment at Brookhaven National Laboratory and has a total uncertainty of 0.54 ppm \cite{e821_result}. Most importantly, there is a difference of greater than three standard deviations between the predicted and measured values. While this discrepancy is suggestive, there is currently an extensive effort underway to improve both the prediction and measurement of $a_\mu$ to determine unequivocally whether there is a discovery of new physics. More information on the theoretical prediction of $a_\mu$, and how that prediction is affected by hadronic cross section measurements can be found in Ref. \cite{gary_proceedings}.

\begin{figure}
\centering
\includegraphics[width=0.75\textwidth]{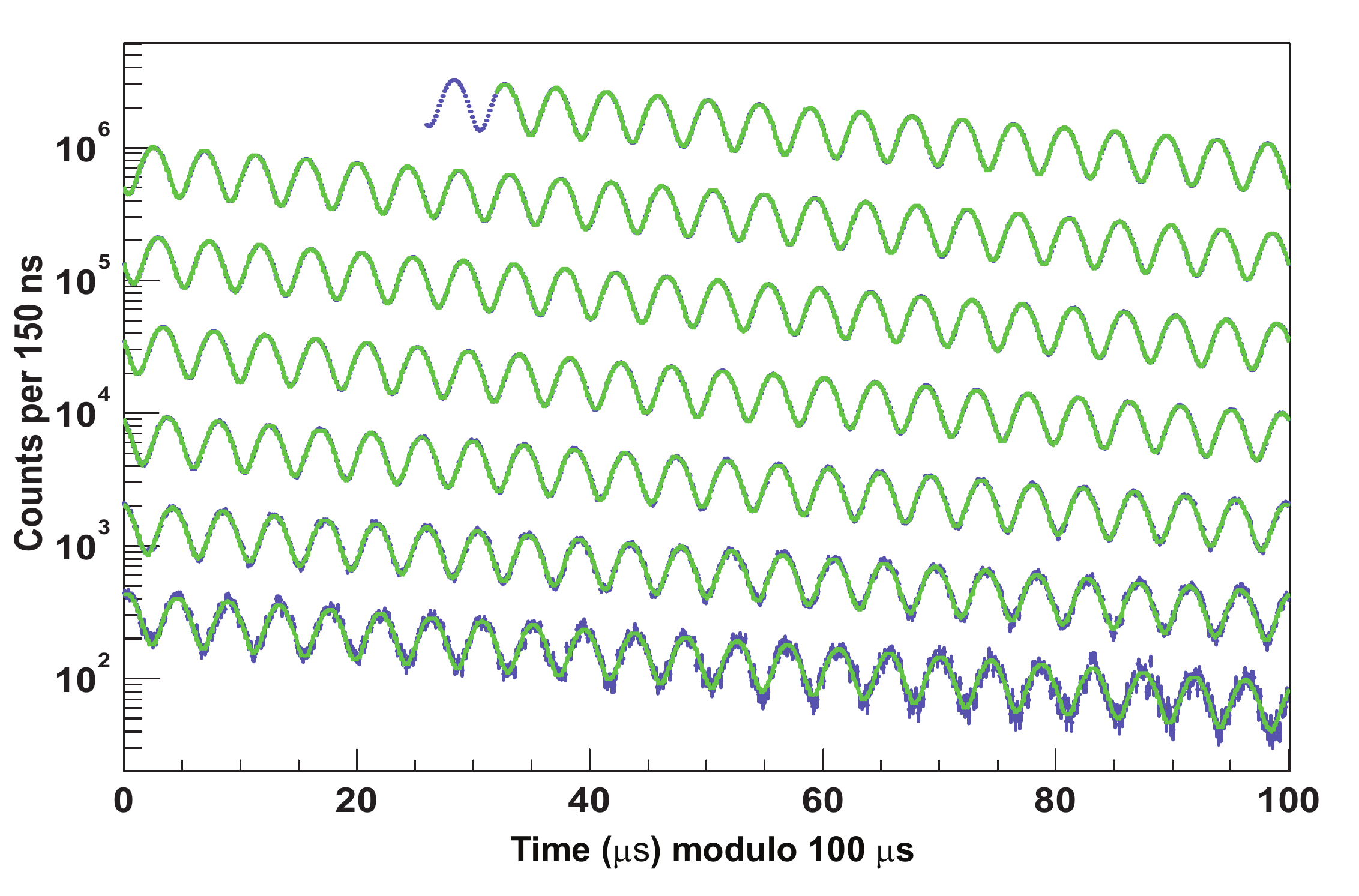}
\caption{The time spectrum of decay positrons above 1.8 GeV/$c^2$ from the E-821 experiment. The modulation in the exponential decay spectrum is at the frequency $\omega_a$, which is directly proportional to $a_\mu$. Figure is from Ref. \cite{e821_result}.}
\label{e821_wiggle}
\end{figure}

The current generation of $a_\mu$ experimental measurements are storage ring experiments. Polarized muons can be produced as a secondary beam. One method is to bombard high energy protons onto a target and to collect the resulting pions. 
Due to the $V-A$ nature of the weak force responsible for the pion decay to muons, the spin of the resulting muons is aligned with the muon momentum. These polarized muons are then inserted into a storage ring magnetic field where the muons circulate at the cyclotron frequency ($\omega_c$). The magnetic field causes the muon spin vector to precess at the Larmor frequency ($\omega_s$). When the muon decays (to a positron\footnote{Both experiments that will be discussed will utilize positive muons, at least for the initial data taking runs.} and two neutrinos), the highest energy positrons are preferentially produced when the muon spin and momentum are aligned. Selecting these highest energy positions imprints a modulation into the muon decay time spectrum with a specific frequency, as shown in Fig. \ref{e821_wiggle}. This frequency ($\omega_a = \omega_s - \omega_c$) is directly proportional to $a_\mu$. 

\section{The E-989 Experiment at Fermilab}

The E-989 experiment at Fermilab represents the next generation of the E-821 experiment \cite{e989_tdr}. During the summer of 2013, the 14-m diameter superconducting coils from the E-821 storage magnet were moved from Brookhaven National Laboratory in New York to Fermilab, near Chicago. Performing the experiment at Fermilab provides a number of advantages, including the ability to produce more muons and to greatly reduce the pion contamination of the muon beam injected into the storage ring, which was a major limiting factor for E-821. The goal of E-989 is to reduce the uncertainty on the measurement of $a_\mu$ to 0.14 ppm, a factor of four improvement relative to the previous result. The experimental collaboration currently consists of approximately one-hundred scientists in thirty institutions around the world \cite{website}.

The equation describing the relationship between $\omega_a$ and the anomalous magnetic moment is 
\begin{equation}
\label{omega_a_eqn}
\overrightarrow{\omega_a} = \frac{e}{mc} \left[ a \overrightarrow{B} - \left(a - \frac{1}{\gamma^2 -1} \right) \overrightarrow{\beta} \times \overrightarrow{E} \right]
\end{equation}

Fermilab will produce muons as a secondary beam, and as a result, the muon beam will have a large emittance. Focusing is necessary to ensure the muons will be stored. While beams typically use magnetic focusing with quadrupole magnets, this would great complicate the measurement of $\omega_a$. Instead, E-989 will utilize electrostatic quadrupoles to focus the muon beam in the storage magnet. By utilizing muons with $\gamma = 29.3$ (corresponding to a muon momentum of 3.09 GeV/$c$, known as the \emph{magic momentum}), the electric field term in Eqn. \ref{omega_a_eqn} will be zero. This results in a simple relationship between $a$, $\omega_a$, and $B$. The key to the experiment, then, is to minimize the uncertainty on the measurement of $\omega_a$ and $B$. 

\begin{figure}
\centering
\includegraphics[width=0.45\textwidth]{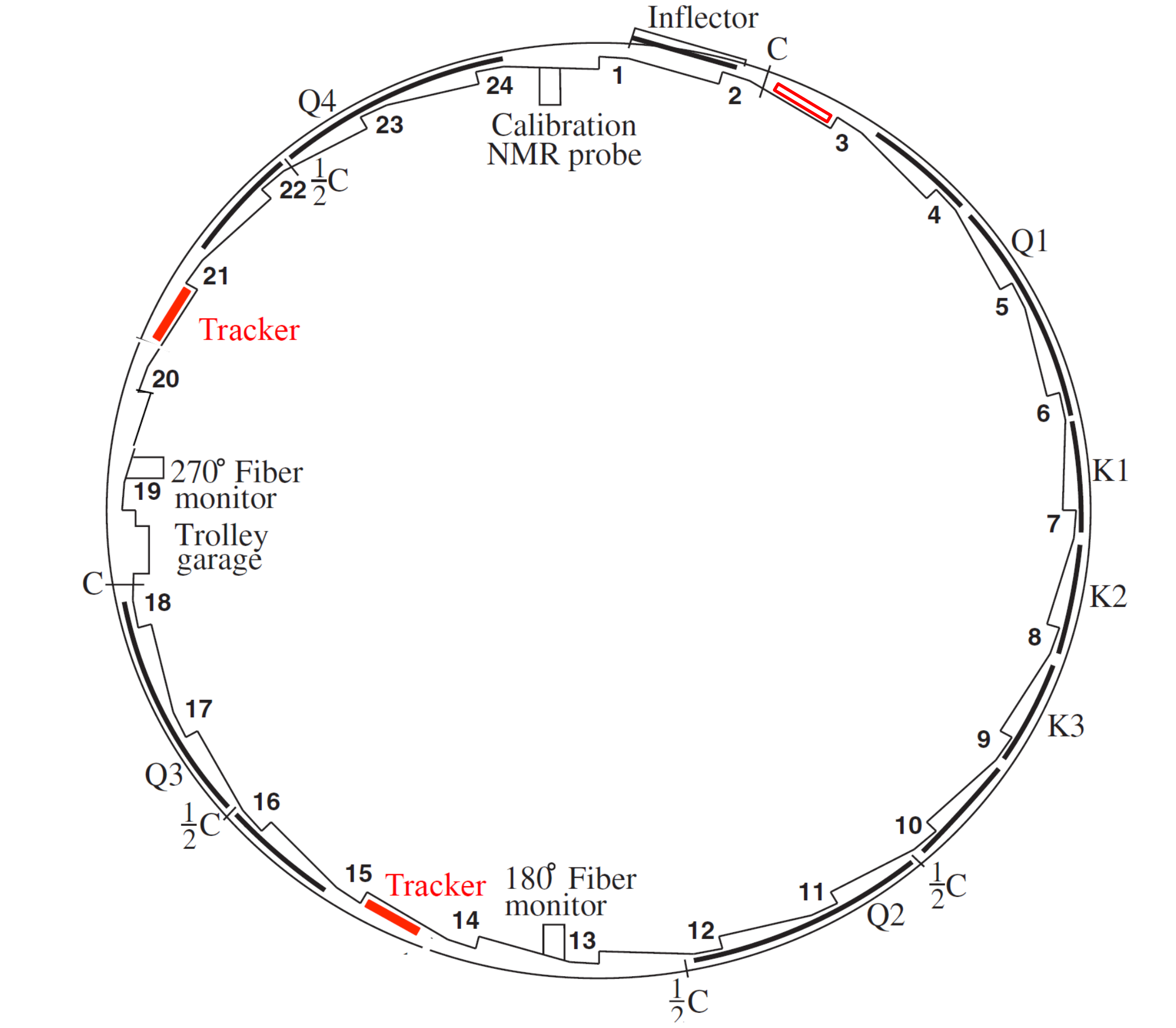}
\includegraphics[width=0.45\textwidth]{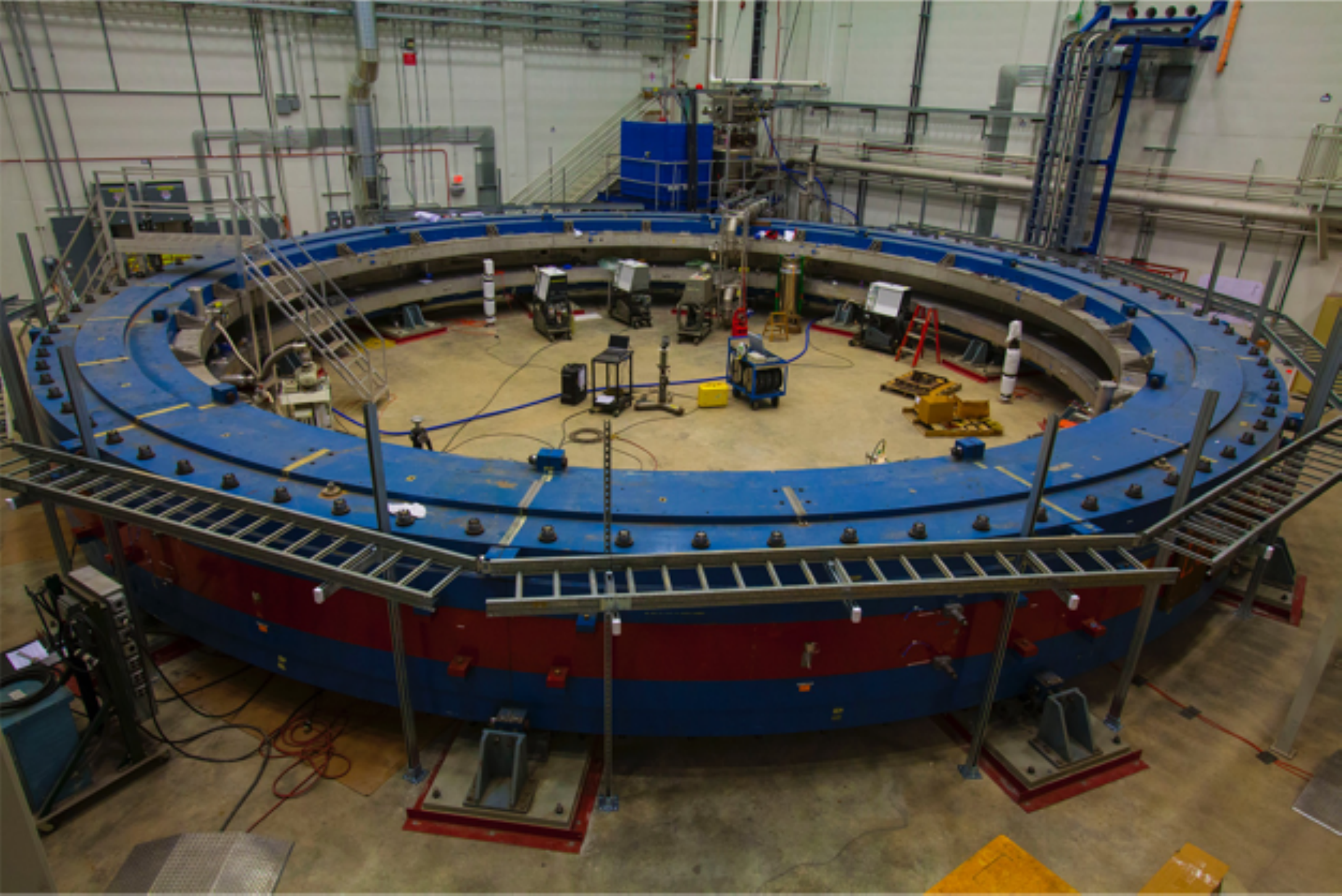}
\caption{Top schematic view of the E-989 storage ring with detector subsystems (left) and current photo of the E-989 storage ring showing the reassembled storage magnet (right).}
\label{e989_detectors}
\end{figure}

The measurement of $\omega_a$ will be performed with a variety of detector subsystems. A schematic of the experiment is shown in Fig. \ref{e989_detectors} (left). Positrons produced from the decay of stored muons will carry only a portion of the initial muon momentum, and hence will be bent into a tighter radius in the storage ring magnetic field. Therefore, twenty-four calorimeter are evenly placed around the inside of the storage ring to measure the energy and arrival time of these decay positions. Each calorimeter station consists of 54 lead fluoride crystals read out by silicon photomultipliers. The segmentation of the calorimeters will reduce the impact of multiple positrons arriving at a single calorimeter closely spaced in time (\emph{pileup}). A dedicated laser calibration system will ensure the calorimeter gain stability. Prototypes of the calorimeter and the laser calibration system have undergone beam tests and have demonstrated the performance needed to achieve the experiment's uncertainty goals \cite{cal_beamtest}. 

In addition to the calorimeters, several other detector subsystems are being constructing. A straw tube tracking system will place eight straw the tracking modules, each with a total of 128 straws in four small-angle stereo layers, in front of three calorimeter stations. This tracking system will measure the trajectory of decay positrons and will allow for the measurement of the beam dynamics parameters of the stored muons. Scintillating fiber detectors (fiber harps) at two locations in the storage ring will be used for a destructive measurement of the properties of the stored muon beam. An entrance counter, based on plastic scintillator, will measure the precise arrival time and intensity of the injected muons. Finally, beam position monitors based on scintillating fibers will be used to measure the precise spatial distribution of the muon bunches as they are injected into the storage ring. 

The measurement of the magnetic field at E-989 will also be improved, compared to E-821. The main method of measuring the 1.45-T field in the storage region relies on approximately 400 nuclear magnetic resonance (NMR) probes. All of these probes are being rebuilt. Most of the NMR probes are mounted in fixed positions on vacuum chambers outside the muon storage region. Seventeen probes are located on a movable trolley which rides on rails and directly measure the magnetic field in the muon storage region. These trolley runs will be performed on a daily to weekly basis. The electronics on the trolley is being completed redesigned. This will provide a much improved data link (allowing for full NMR wave forms to be stored) and a optical bar code system to precisely measure the position of the trolley. The trolley and fixed NMR probes allow for a relative calibration of the magnetic field experienced by the muons around the ring. In addition, there is another water-based NMR probe that will provide the absolute calibration. It is currently planned to use this absolute calibration probe to cross-calibrate with the proposed E-34 experiment at J-PARC.\footnote{It is also planned to use this absolute calibration probe to cross-calibrate the magnetic field in the planned MuSEUM experiment at J-PARC. This experiment will measure the hyperfine splitting of muonium and provides an important input to the $a_\mu$ measurements.} Work is also ongoing to build another absolute calibration probe based on Helium-3. This second absolute calibration probe will provide an important cross-check on many systematic uncertainties present in the magnetic field measurement. 

The upgrades to the E-989 experiment will provide a substantial reduction in the uncertainty of the $a_\mu$ measurement, as compared with the final E-821 result. The detector improvements will reduce the systematic uncertainty on the $\omega_a$ measurement from 0.17 ppm in E-821 to 0.07 ppm. The improvements to the magnetic field measurement will reduce the systematic uncertainty from the 0.18 ppm obtained in E-821 to 0.07 ppm. A planned two-year run will provide enough muons (on the order of $10^{11}$ in the final $\omega_a$ fit) to have a statistical uncertainty of 0.01 ppm. The current status of the experiment is that the storage magnet has been reassembled, cooled, and powered. Next, follows a six to nine month period of course mechanical shimming of the magnetic field. The detector subsystems will be installed and commissioned starting in summer of 2016. First beam is expected in fall 2017. 

\section{The E-34 Experiment at J-PARC}

Another planned experiment to measure $a_\mu$ is the E-34 experiment at J-PARC in Japan \cite{e34_ref}. While still a storage ring experiment, it takes a very different approach to the E-989 experiment at Fermilab. E-989 operated at the muon magic momentum and utilized electrostatic quadrupole focusing to simplify the extraction of $a_\mu$ from Eqn. \ref{omega_a_eqn}. E-34 will utilize ultra-cold muons, eliminating the need for focusing. This results in an experiment that will have very different systematic uncertainties, and hence will be very complementary to E-989. 

\begin{figure}
\centering
\includegraphics[width=0.75\textwidth]{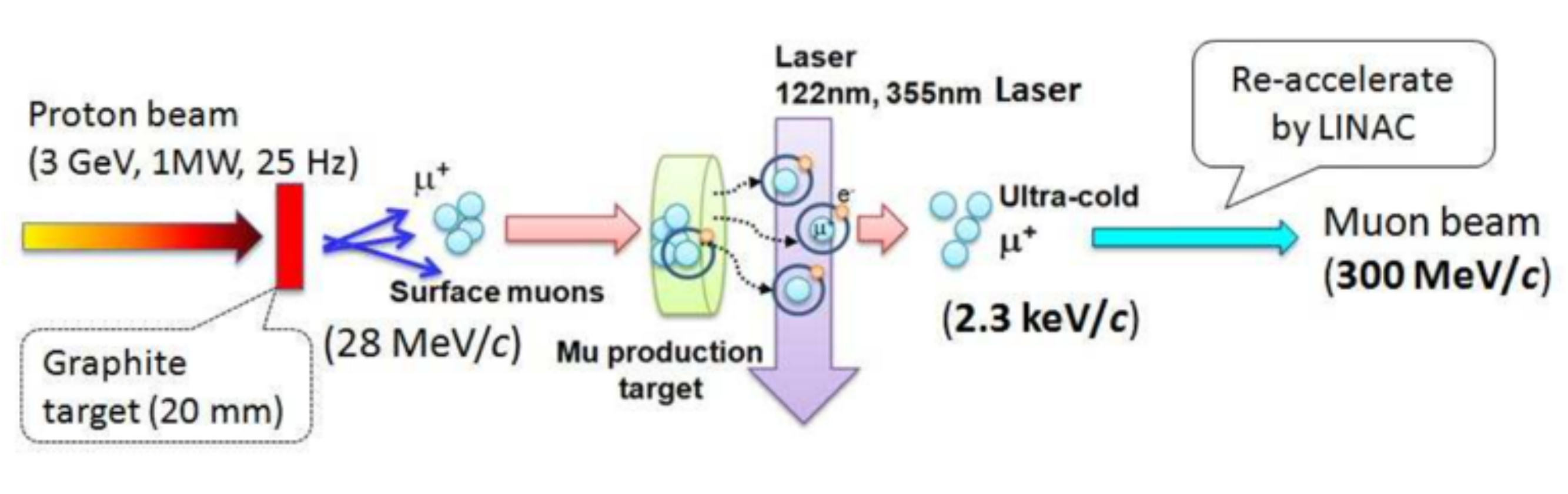}
\caption{Steps in ultra-cold muon production at J-PARC.}
\label{muon_production}
\end{figure}

The key to E-34 is the production of ultra-cold muons. If the emittance of the produced muon beam is small enough, then there is no need for strong focusing to ensure that most muons will be stored in the storage ring. The production of ultra-cold muons at J-PARC, the steps of which are shown in Fig. \ref{muon_production}, will begin with 3 GeV/$c^2$ protons striking a graphite target. This will produce pions, some of which will stop near the surface of the target. The low-momentum polarized muons produced from the pion decays (known as \emph{surface muons}) will then drift and will capture electrons in a second target to form muonium. Some of these muonium atoms will drift off the room-temperature target and then will be ionized with two separate laser pulses. This intermediate muonium will result in a dilution of the muon polarization (down to 0.5) as not all of the muonium spin states will preserve the initial muon polarization. The muons are then reaccelerated to a momentum of 300 MeV/$c$ before being injected into the storage magnet. Before reacceleration, the muons will have momentum on the order of a few keV/$c$, resulting in an extremely small transverse beam emittance after acceleration. The production rate of ultra-cold muons is currently a limiting factor. However, recent experimental efforts with a muonium target of silica aerogel with laser-ablated micro-channels has shown a significant improvements over previous results \cite{triumf_muonium}. Currently, the predicted rate of ultra-cold muons into the experiment's storage ring is $0.2 \times 10^6$/s. While less than the projected rate of $1 \times 10^6$/s in the experimental proposal, it is likely that near future efforts can continue to increase the ultra-cold muon production rate. 

\begin{figure}
\centering
\includegraphics[width=0.45\textwidth]{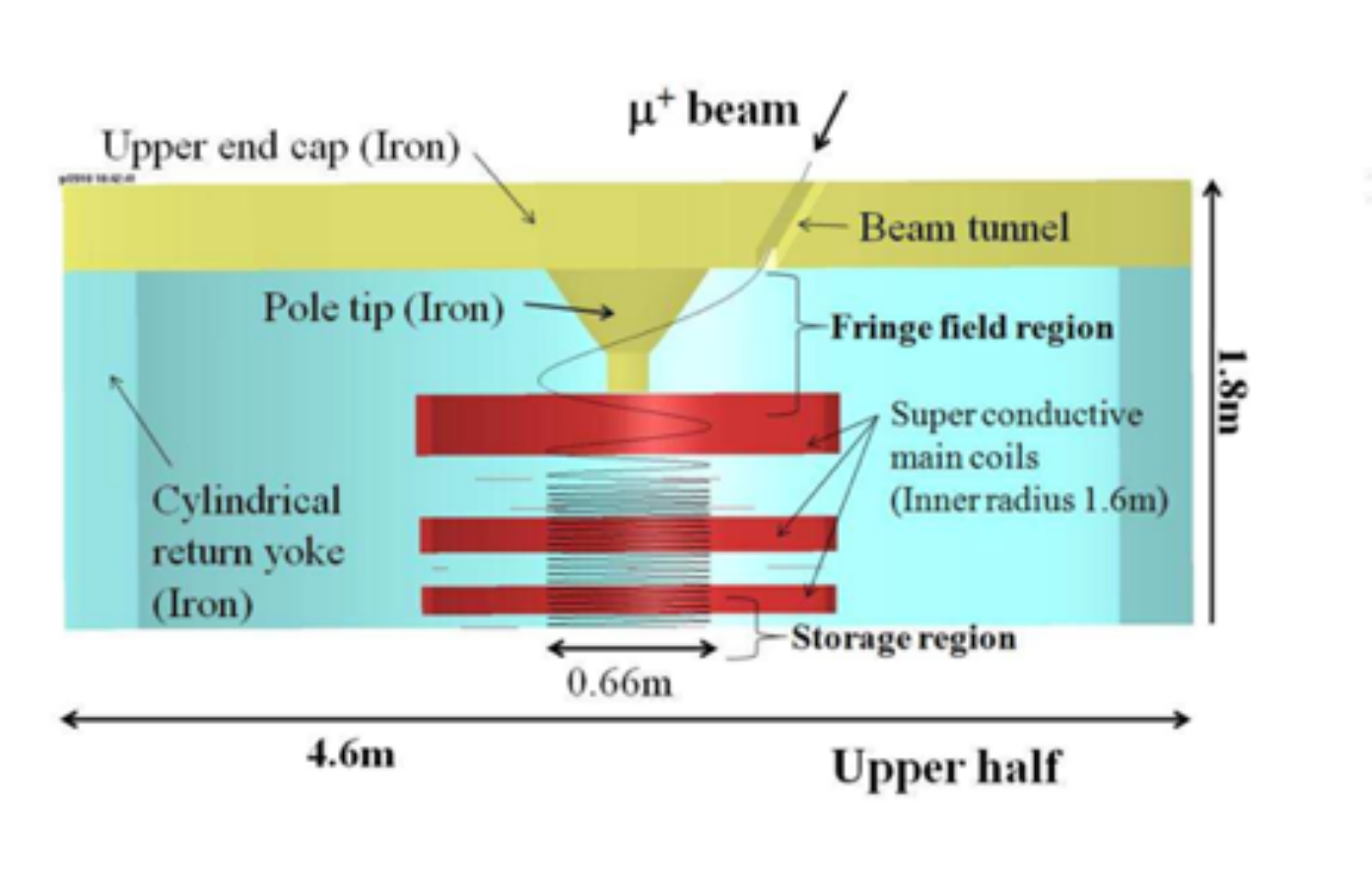}
\includegraphics[width=0.45\textwidth]{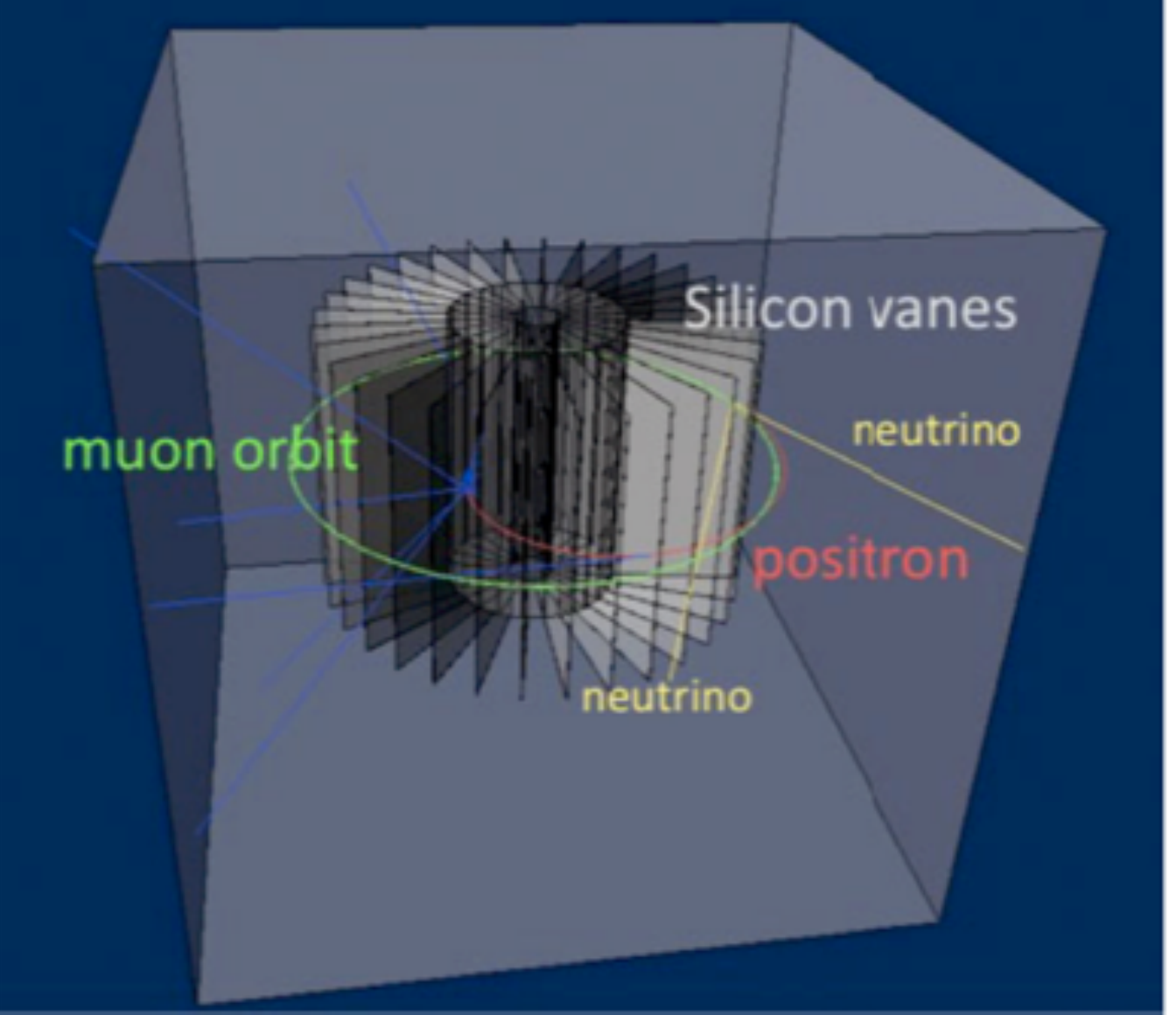}
\caption{The E-34 storage magnet (left) and the silicon sensors (right).}
\label{e34_magnet}
\end{figure}

Since the ultra-cold muons are produced with relatively low momentum (300 MeV/$c$), the storage ring can be fairly compact. E-34 will utilize a superconducting 66-cm diameter, 3.0-T solenoid magnet to store the muons. The small size of the magnet should make it much easier to produce a uniform magnetic field, thereby reducing the systematic uncertainty on the magnetic field measurement. The muons are injected at an angle with respect to the horizontal, then pulsed electromagnets will be used to kick the muons on the the horizontal storage orbit, as shown in Fig. \ref{e34_magnet} (left). After the decay of the muons, the produced positrons will be detected by 98 planes of silicon sensors located in the center of the solenoid, inside the muon storage region. These silicon sensors cover 6 cm in radius and 12 cm height. As in E-989, selecting the highest energy positrons will cause the decay time spectrum to be modulated by the $\omega_a$ frequency. 

The E-34 experiment is anticipating an initial one year run, which based on current estimates, should produce a statistical uncertainty on $a_\mu$ of 0.4 ppm. (Compared to a projected statistical uncertainty of 0.1 ppm for a two-year run of E-989.) Estimations of systematic uncertainties are still underway. The exact time scale of the experiment is still uncertain, but once final approval and funding is secured, the experiment could be ready for data taking in a few years time. Current estimates have data taking beginning in 2019. 

\section{Conculusion}

The history of improvements in the prediction and measurement of the magnetic moments of fundamental particles has paralleled the development of the standard model of particle physics and has been an important factor in our improving understanding of particle physics. Most recently, the sub-ppm measurement and prediction of the anomalous magnetic moment of the muon show a tantalizing disagreement which could be an indication of the presence of new particles or forces. Significant efforts are now underway to improve the precision of both the prediction and the measurement. Two new experimental efforts are underway. The E-989 experiment at Fermilab is a significant upgrade and evolution of the E-821 experiment at Brookhaven National Laboratory. Improvements to the detectors, field measurements, and beam lines are anticipated to result in a total uncertainty on the $a_\mu$ measurement of 0.14 ppm, a factor of four improvement on E-821. The two-year data taking run is scheduled to begin in 2017. The E-34 experiment proposed at J-PARC utilizes a significantly different approach, relying on a low-emittance ultra-cold muon beam and high precision tracking in a small storage volume. Significant challenges remain, but data taking could begin as early as 2019. These two experiments have the potential to definitively probe whether new physics is present in $a_\mu$.

I would like to thank Fermi National Accelerator Laboratory (Fermilab), the Japan Proton Accelerator Research Complex (J-PARC), the Japanese Ministry of Education, Culture, Sports, Science, and Technology (MEXT), the United States Department of Energy (DOE), the United States National Science Foundation (NSF), and the conference organizers for an excellent and enlightening event.


\end{document}